\documentclass[11pt]{article}

\usepackage{graphicx}
\usepackage{amsmath, amsthm, amssymb}
\usepackage{algorithm}
\usepackage{algorithmic}
\usepackage{subfigure}
\usepackage{comment}
\usepackage{url}
\usepackage{pgf}

\input xy
\xyoption{all}
\newcommand{\edge}[1]{\ar@{-}[#1]}
\newcommand{\node}{*+[o][F-]{ }}

\newcommand{\Z}{\mathbb{Z}}

\newcommand{\ket}[1]{| #1 \rangle}

\newcommand{\be}{\begin{equation}}
\newcommand{\ee}{\end{equation}}
\newcommand{\bea}{\begin{eqnarray}}
\newcommand{\eea}{\end{eqnarray}}
\newcommand{\bes}{\begin{equation*}}
\newcommand{\ees}{\end{equation*}}
\newcommand{\beas}{\begin{eqnarray*}}
\newcommand{\eeas}{\end{eqnarray*}}

\newtheorem{thm}{Theorem}[section]
\newtheorem*{thm*}{Theorem}

\newtheorem{lem}[thm]{Lemma}
\newtheorem*{lem*}{Lemma}

\begin{document}

\date{\today}

\title{Quantum search of partially ordered sets}

\author{Ashley Montanaro\footnote{montanar@cs.bris.ac.uk}\\ \\ {\small Department of Computer Science, University of Bristol,}\\{\small Woodland Road, Bristol, BS8 1UB, UK.} }

\maketitle

\begin{abstract}
We investigate the generalisation of quantum search of unstructured and totally ordered sets to search of partially ordered sets (posets). Two models for poset search are considered. In both models, we show that quantum algorithms can achieve at most a quadratic improvement in query complexity over classical algorithms, up to logarithmic factors; we also give quantum algorithms that almost achieve this optimal reduction in complexity. In one model, we give an improved quantum algorithm for searching forest-like posets; in the other, we give an optimal $O(\sqrt{m})$-query quantum algorithm for searching posets derived from $m \times m$ arrays sorted by rows and columns. This leads to a quantum algorithm that finds the intersection of two sorted lists of $n$ integers in $O(\sqrt{n})$ time, which is optimal.
\end{abstract}


\section{Introduction}

Searching for an object in a set of objects that obey some structure is a fundamental task in computer science. The archetypal example of such a task is finding an integer in a sorted list containing $n$ elements; in this case, binary search can find the marked integer in $O(\log n)$ steps. At the other extreme, any (classical) search algorithm requires $\Omega(n)$ steps to search a completely unsorted $n$-element list. It is thus of interest to find a framework for search problems that encompasses both of these structures, and interpolates between them.

One approach is to consider the task of searching a partially ordered set ({\em poset}). Recall that a partial order on a set $S$ is a relation $\le$ such that, for $a,b,c \in S$, $a \le a$, $(a \le b) \wedge (b \le a) \Rightarrow a=b$, and $(a \le b) \wedge (b \le c) \Rightarrow a \le c$. We define the relation $<$ in the obvious way: $(a<b) \Leftrightarrow (a \le b) \wedge (a \neq b)$. For any two elements $a,b$, either $a \le b$, $b \le a$, or $a$ and $b$ are incomparable, $a \nleq b$. We say that a set is totally ordered if none of its elements are incomparable, and unstructured if all of its elements are incomparable.

There are two natural ways to model poset search. In the first model (introduced by Linial and Saks \cite{linial}, and called the {\em concrete} model here), we consider the partial order on $S$ to represent constraints on the structure of an unknown {\em totally} ordered set, identified with the integers. That is, each element $s\in S$ stores an integer $x=S[s]$, which is returned by a query to the element $s$. The constraint following from the partial order on $S$ is that if $s \le t$ for some $s,t \in S$, then $S[s] \le S[t]$. The goal is to find the location at which a (known) arbitrary integer $a$ is stored, or to output that $a$ is not stored in $S$, using the minimum number of queries to elements of $S$. We will usually assume that the integers stored in $S$ are all distinct.

Alternatively, in the second model (introduced by Ben-Asher, Farchi and Newman \cite{benasher}, and called the {\em abstract} model here), the goal is to search for an unknown ``marked'' element $a \in S$, using the minimum number of queries to an oracle, which operates in the following way. On input of an element $x \in S$, the oracle returns one of $\{<,=,\nleq\}$. The first two possibilities are returned when $a<x$ and $a=x$, respectively, and the third is returned when either $x<a$ or $x$ and $a$ are incomparable.

We sometimes mention an extension of the search problem to a scenario where multiple different answers are permissible. This extension is different for the two models: in the abstract model, we consider there to be multiple marked elements in the set to be searched, with the goal being to output any of these elements. In the concrete model, the analogous scenario is allowing the possibility for the set to store duplicate integers, i.e.\ allowing there to exist $s,t\neq s$ such that $S[s]=S[t]$.

To sum up, in the concrete model we know what we are searching for, but not where to find it; in the abstract model, we do not know what we are searching for, but we can perform powerful queries that narrow down the search space to find it.

This paper is concerned with quantum search of posets in both of these models, and in particular with minimising the number of queries to the set required to find the desired element. It is well-known that Grover's algorithm \cite{grover} can find the marked element in an unstructured $n$-element set using $O(\sqrt{n})$ quantum queries, thus gaining a quadratic advantage over classical computation, and that this reduction is optimal. However, no advantage beyond a constant factor may be achieved for quantum search of a totally ordered set \cite{hoyer}.

We then have several questions, motivated by these two examples. Can we find interesting quantum algorithms for search of general posets? Could a reduction in queries of more than the quadratic factor given by Grover's algorithm be achieved by such an algorithm, or even an exponential reduction? And what is the structure (or otherwise) of the posets for which a quantum computer can gain an asymptotic advantage over classical computation?


\subsection{New results}

Our first result is that, in both the abstract and concrete models, quantum algorithms can achieve no more than a quadratic reduction (up to a logarithmic factor) in the number of oracle queries to find a marked element. The lower bounds in the two models seem to need different proof techniques: the bound in the abstract model follows from a reduction to the oracle identification problem of Ambainis et al \cite{ambainis}, whereas we use structural properties of posets to derive the lower bound in the concrete model.

We give general upper bounds that match these lower bounds up to logarithmic factors. In the abstract model, the upper bound follows from an algorithm of Atici and Servedio \cite{atici}. In the concrete model, we give a new and almost optimal quantum algorithm that follows from Dilworth's Theorem \cite{dilworth} on the decomposition of posets into ordered components.

These general results can be summarised as the following theorem.

\begin{thm}
Let $S$ be an $n$-element poset, and let $D(S)$ and $Q_2(S)$ be the number of queries required for an exact classical or bounded-error quantum (respectively) algorithm to find the marked element in $S$, in either of the two models discussed above. Then
\beas
D(S) &=& O(Q_2(S)^2\log n)\\
Q_2(S) &=& \left\{ \begin{array}{lr} O(\sqrt{D(S)}\log n \sqrt{\log \log n}) & \mbox{(abstract model)}\\
O(\sqrt{D(S)}\log n) & \mbox{(concrete model)}\end{array}\right.
\eeas
\end{thm}

In both models, we give explicit quantum algorithms for searching specific poset structures. In the abstract model, we give a simple (and nearly optimal) algorithm for searching a class of forest-like posets. For an unstructured set, the algorithm reduces to Grover search, whereas for a totally ordered set it reduces to binary search.

In the concrete model, we give an asymptotically optimal quantum algorithm for searching posets that are derived from 2-dimensional arrays of distinct integers sorted by rows and columns. This gives rise to an optimal quantum algorithm for an apparently unrelated problem: finding the intersection of two sorted lists. Given two lists of at most $n$ integers in increasing order, the algorithm can find an element that appears in both lists in $O(\sqrt{n})$ time, improving on a previous algorithm of Buhrman et al \cite{buhrman}, which achieved a time complexity of $O(\sqrt{n}c^{\log^*n})$ for some constant $c$.


\subsection{Previous work}

Classically, the question of searching partially ordered sets seems to have first been considered by Linial and Saks \cite{linial,linial2}, who characterised the query complexity of searching posets in their concrete model. They showed that this complexity depends solely (up to constant factors) on the number of ideals of the poset, where an ideal of $S$ is a subset $T\subseteq S$ such that $(x\in T)\wedge(y<x) \Rightarrow (y\in T)$. In particular, they give lower and upper bounds on the complexity of searching for a marked element in an array sorted by rows and columns, and the $d$-dimensional generalisation thereof.

Ben-Asher, Farchi and Newman \cite{benasher} introduced the abstract model, and gave an algorithm to find the optimal search strategy in this model for a class of tree-like posets. In this model, it is interesting to note that the problem of determining an optimal search strategy for arbitrary posets is NP-hard, whereas the same question restricted to trees is soluble in polynomial time \cite{carmo}. In fact, Onak and Parys have recently obtained an $O(n^3 \log n)$-time algorithm for finding this strategy \cite{onak}, and also point out that this model is similar to a model of search in graphs, where one queries an edge and is returned the closest endpoint of that edge to the marked element. It was already known that near-optimal search strategies for almost all posets can be produced efficiently \cite{carmo}.

In the case of quantum search, tight upper and lower bounds on query complexity are known for search of unstructured sets \cite{grover,brassard,zalka}. An asymptotically tight lower bound is known for search of totally ordered sets \cite{ambainis,hoyer}. We will also make use of related results by Aaronson and Ambainis on spatial quantum search \cite{aaronson2}.


\section{Preliminaries}


\subsection{Quantum query algorithms}

In this work, the measure used of the complexity of searching a poset $S$ is usually the number of queries to $S$ required to find the marked element, or report that none exists, rather than the time required for the search (see Section \ref{sec:conclusions} for a brief discussion of this point).

We will assume familiarity with quantum computation, and will use the standard model of quantum query complexity. In this model, a $t$-query quantum algorithm is a sequence of unitary transformations $U_t O_a U_{t-1} O_a \cdots O_a \ket{\psi}$, where we alternate between ``expensive'' oracle queries that may depend on an unknown entity $a$, and ``free'' arbitrary unitary operations that do not, with the aim being to minimise the number of oracle queries. The oracle $O_a$ is usually taken to be a unitary operator that operates on an $n$-dimensional input register $\ket{x}$ and $d$-dimensional output register $\ket{y}$, and encodes an arbitrary function $f_a(x):\Z_n \mapsto \Z_d$ as follows: $O_a\ket{x}\ket{y}=\ket{x}\ket{y+f_a(x)}$, where addition is taken modulo $d$.

In the abstract model, we require an oracle $f_a(x)$ that returns something from the set $\{<,=,\nleq\}$, according to whether the unknown marked element $a<x$, $a=x$ or $a\nleq x$. However, it will be convenient to instead use a Boolean oracle by adding a parameter $z\in\{0,1\}$ to give an oracle function $f_a(x,z)$, which acts as follows. $f_a(x,0)=1$ if $a\le x$, and 0 otherwise. $f_a(x,1)=1$ if $x=a$, and 0 otherwise. It is clear that a query to $f_a(x)$ is sufficient to simulate a query to $f_a(x,z)$, and querying $f_a(x,0)$ and $f_a(x,1)$ is sufficient to simulate $f_a(x)$. The query complexity in the two-parameter model may thus only differ by a factor of at most 2 from the one-parameter model. The model can be extended to allowing more than one marked element in an obvious way, by parametrising the oracle with a set of marked elements $A$; then $f_A(x,0)=1$ if there exists $a \in A$ with $a \le x$.

The concrete model is more straightforward; here, the oracle depends only on the integers stored in the set $S$, and an oracle query to an element $x$ simply returns the integer stored at the element $x$, i.e.\ $S[x]$. We usually assume that, for all $x \neq y$, $S[x] \neq S[y]$.

$D(S)$ will denote the worst-case exact classical decision tree complexity of searching for a single marked element in the poset $S$, and $Q_E(S)$ the equivalent quantum query complexity. $Q_2(S)$ is the quantum query complexity where we are allowed to err with probability $\le 1/3$ (the ``2'' refers to 2-sided bounded error). Motivated by binary search, our notion of a poset $S$ that allows ``efficient'' search is one where the marked element can be found using a number of queries that is polylogarithmic in $|S|=n$. All logarithms will be taken to base 2.

We will make frequent use of an exact variant of Grover's quantum search algorithm \cite{grover}.

\begin{thm}
\label{thm:exactGrover}
{\bf(Exact Grover search [e.g.\ \cite{brassard}, \cite{long}])}\\
Let $S$ be an unstructured set of $n$ elements containing either one marked element, or no marked elements. Then there exists an exact quantum algorithm which outputs the marked element, or that no such element exists, using $O(\sqrt{n})$ queries to the set.
\end{thm}


\subsection{Posets}

We will use standard terminology relating to posets. A {\em chain} in a poset $S$ is a subset $T\subseteq S$, all of whose elements are comparable. Conversely, an {\em antichain} is a subset whose elements are all incomparable. The {\em height} $h(S)$ and {\em width} $w(S)$ of a poset $S$ are the size of the largest chain and antichain in $S$, respectively. A {\em subset} of a poset $S$ is a subset of the elements in $S$ that preserves the order relations; conversely, an extension of $S$ preserves the elements but may add new order relations. A {\em section} of $S$ is a subset $T\subseteq S$ such that $(x\in T) \wedge (z\in T) \wedge (x<y<z) \Rightarrow y \in T$. A {\em maximal element} of $S$ is an element $x \in S$ such that for all $y \in S$, $y \ngtr x$.

\begin{figure}[htp]
\centering
\begin{tabular}{c@{\hspace{2cm}}c@{\hspace{2cm}}c}
\xymatrix@=10pt{
\node\edge{d}\\
\node\edge{d}\\
\node\edge{d}\\
\node
}
\vspace{10pt}
&
\put(3,-25){
\xymatrix@=10pt{
\node & \node & \node & \node
}}
&
\put(-20,-10){
\xymatrix@=10pt{
& & & \node & & & \\
& \node\edge{urr} & & & & \node\edge{ull} & \\
\node\edge{ur} & & \node\edge{ul} & & \node\edge{ur} & & \node\edge{ul}
}}
\\
Totally ordered set & Unstructured set & Tree-like poset
\end{tabular}
\caption{Hasse diagrams of some posets}
\label{fig:hasse}
\end{figure}
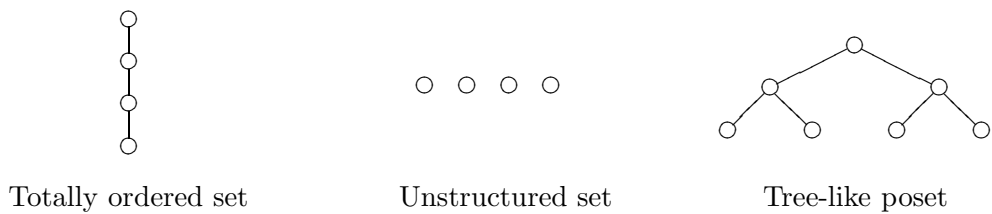

A poset can be represented graphically by its {\em Hasse diagram}. A Hasse diagram for $S$ is an undirected graph $G$ whose vertices are labelled by the elements of $S$. We say that $b$ covers $a$ if $b>a$ and there does not exist $c \in S$ such that $a < c < b$. For each pair of vertices $a,b$, if $a$ covers $b$ then the vertex corresponding to $a$ in the Hasse diagram is connected to, and positioned higher than, that corresponding to $b$. Figure \ref{fig:hasse} gives the Hasse diagrams of some example posets.

A poset $S$ is said to be tree-like (forest-like) if its Hasse diagram is a tree (forest) rooted at the maximal element(s) of $S$.


\section{The abstract model}

In this section, we consider the problem of searching posets in the model studied by Ben-Asher, Farchi and Newman \cite{benasher}, where a query to an element of a poset $S$ returns information about its relationship to the unknown marked element with respect to the partial order on $S$.


\subsection{Overall relationships}

In this model, we can immediately relate quantum and classical search using a reduction to the oracle identification problem, which was originally introduced (in the context of quantum computation) by Ambainis et al \cite{ambainis}, and related to computational learning theory by Servedio and Gortler \cite{servedio}. In this problem, we are given as an oracle an unknown $m$-bit Boolean function $f$ picked from a known set of functions $S$, and we must identify $f$ with the minimum number of queries to the oracle (Servedio and Gortler refer to this as {\em exactly learning} $f$).

Servedio and Gortler have shown \cite{servedio} that the quantum and classical query complexities of this task are closely related, and both depend on a parameter which we call $\gamma^S$\footnote{This is Servedio and Gortler's $\hat{\gamma}^C$.}, which is informally defined as the minimum fraction of the functions in $S$ which a classical algorithm can be certain of removing from consideration with a query to $f$. To be precise, let $S'$ be a subset of $S$, and let $S'_{a,b}$ be the subset of those functions in $S'$ that take value $b$ on input $a$. Then
\be \gamma^S = \min_{S' \subseteq S, |S'|\ge 2} \max_{a\in\{0,1\}^m} \min_{b\in\{0,1\}} \frac{|S'_{a,b}|}{|S'|}\ee
The main result of \cite{servedio} may be stated as:
\begin{thm}
\label{thm:servedio}
{\bf \cite{servedio}}\\
Let $S$ be a set of Boolean functions on $m$ bits. Then the quantum query complexity $Q$ of exactly learning a function from $S$, with a bounded probability of error, obeys the following lower bounds.
\be Q = \Omega\left(\frac{1}{\sqrt{\gamma^S}}\right),\, Q = \Omega\left( \frac{\log |S|}{m} \right) \ee
Also, the deterministic classical query complexity $C$ of the same task obeys the following upper bound.
\be C = O\left(\frac{\log |S|}{\gamma^S}\right) \ee
Quantum and classical query complexities are thus related by $C=O(mQ^3)$.
\end{thm}
The classical algorithm that achieves this query complexity is quite straightforward, simply consisting of querying the unknown function at the input that, given an adversarial response, reduces the size of the set of remaining possible functions by the largest possible amount.

We now make a connection between the poset search problem and oracle identification. Given a poset, the oracle associated with each possible marked element $a$ is a two-parameter Boolean function $f_a(x,z)$. Distinguishing between these functions is exactly equivalent to finding the hidden $a$. Thus, in order to find the marked element in an $n$-element poset, we need to distinguish $n$ Boolean functions on $\lceil \log n + 1\rceil$ bits. Theorem \ref{thm:servedio} immediately gives the following result.
\begin{thm}
Let $S$ be an $n$-element poset. Then $D(S)=O(\log n\,Q_2(S)^2)$.
\end{thm}
A quadratic reduction in queries is thus the best that can be obtained using a quantum algorithm, up to a logarithmic factor. We now turn to upper bounds on quantum query complexity. There is a straightforward general upper bound of $O(\sqrt{n})$ oracle queries for any poset. This can be seen by noting that, if the oracle $f_a(x,z)$ is queried only with $z=1$, the problem reduces to unstructured search, so Grover's algorithm \cite{grover} can be used.

Less trivially, Atici and Servedio \cite{atici} have given a quantum algorithm for exact learning that can be seen as an analogue of the classical algorithm mentioned in Theorem \ref{thm:servedio}. This algorithm immediately applies to poset search, and moreover is efficient (runs in time polynomial in $n$).
\begin{thm}
\label{abs:upperBound}
{\bf \cite{atici}}
Let $S$ be an $n$-element poset. Then
\be Q_2(S) = O\left(\frac{\log n \log \log n}{\sqrt{\gamma^S}}\right) \ee
\end{thm}
This upper bound can actually be improved to $Q_2(S) = O\left(\log n \sqrt{\log \log n}/\sqrt{\gamma^S}\right)$. The reason is that the $O(\log \log n)$ factor in Atici and Servedio's algorithm comes from perfoming $O(\log \log n)$ rounds of classical probability amplification, which can be replaced by the use of a quantum algorithm of Buhrman et al \cite{buhrman2} that performs efficient amplitude amplification to small error probabilities.

In summary, it can be seen that the quantum and classical query complexities of this search problem are completely determined (up to logarithmic factors) by this parameter $\gamma^S$. However, it is unclear whether the extension to searching for multiple marked elements has a similar reduction to the oracle identification problem, and whether a suitable adaptation of Atici and Servedio's algorithm can be applied in this case.

Finally, note that one might consider a more powerful variant of search in this model, where the oracle $f_a(x)$ is extended to return $>$ if the marked element $a>x$ (so the four possible results are ``$<$'', ``$=$'', ``$>$'' and ``incomparable''). The reduction to the oracle identification problem clearly still holds for this variant, so the results in this section go through unchanged.


\subsection{Search in forest-like posets}

We say a poset is forest-like if every element in the poset is covered by at most one other element (an example of such a poset is shown in Figure \ref{fig:hasse}). Classically, forest-like posets have proven to be easier to analyse; indeed, algorithms exist \cite{benasher,onak} for computing the optimal classical decision tree to search these posets in polynomial time, whereas the same problem is NP-hard for general posets \cite{carmo}. In this section, we present an exact quantum algorithm for searching a forest-like poset $S$ using $O\left(\log n/\sqrt{\gamma^S}\right)$ queries, improving on the previously mentioned bounded-error $O\left(\log n \sqrt{\log \log n}/\sqrt{\gamma^S}\right)$-query algorithm \cite{atici}. Our algorithm improves on Atici and Servedio's in other ways too: firstly, it reduces to an asymptotically optimal algorithm in the case of search of unstructured and totally ordered sets; secondly, it can easily be extended to searching for multiple marked elements, with a small penalty in query complexity.

We first consider the case of a single marked element. The principles behind the algorithm that we will describe are very similar to those underlying Atici and Servedio's. Throughout the algorithm, a subset of possible places that the marked element could be is maintained. We will show that one use of Grover's algorithm over a set $G$ of size at most $1/\gamma^S$ can be used to reduce the size of this subset by at least half, so $\log n$ repetitions suffice to find the marked element. Crucially, for forest-like posets where there is a single marked element, this use of Grover's algorithm can be made exact (Theorem \ref{thm:exactGrover}), thus avoiding the need for some number of repetitions to achieve a suitable reduction in the error probability.

The algorithm is explicitly stated as Algorithm \ref{alg:forest} below. It uses a function \texttt{centralElement} which requires some explanation. Define the weight $wt(v)$ of an element $v\in S$ as $wt(v)=|\{x:(x\in S) \wedge (x\le v)\}|$. Then \texttt{centralElement}($S$) returns the element $v \in S$ such that $wt(v)$ is maximised, given that $wt(v)\le \lceil |S|/2 \rceil$. Such an element will clearly always exist. \texttt{siblings}($x$) returns the set of elements of $S$ that are covered by the single element that covers $x$.

\renewcommand{\algorithmicrequire}{\textbf{Input:}}
\renewcommand{\algorithmicensure}{\textbf{Output:}}

\begin{algorithm}[ht]
\setlength{\parskip}{0pt}
\begin{algorithmic}
\REQUIRE{Forest-like poset $S$ containing $n$ elements}
\ENSURE{Marked element, or ``not found''}
\STATE $T\leftarrow S$;
\WHILE{$|T|>1$}
\STATE $x \leftarrow$ \texttt{centralElement}($T$);
\IF{\normalfont $x$ is a maximal element of $T$}
\STATE $G=\{y:y\,\mbox{is a maximal element of $T$}\}$;
\ELSE
\STATE $G=\{y:y \in \texttt{siblings}(x)\}$;
\ENDIF
\STATE $y \leftarrow$ result of exact Grover search on $G$;
\IF{\normalfont result is ``not found''}
\STATE $T \leftarrow T\,\backslash\{z:\exists y' \in G, z \le y'\}$;
\ELSE
\STATE $T \leftarrow \{z: z \le y\}$;
\ENDIF

\ENDWHILE
\IF{$|T|$=1}
\STATE {\bf return} single element in $T$;
\ELSE
\STATE {\bf return} not found;
\ENDIF
\end{algorithmic}
\caption{Search algorithm for forest-like posets}
\label{alg:forest}
\end{algorithm}

We will now prove an upper bound on the query complexity of Algorithm \ref{alg:forest}, via a couple of preparatory lemmas.

\begin{lem}
\label{lem:reduceByHalf}
In each iteration of the loop, the total weight of the nodes in $G$ is at least $|T|/2$.
\end{lem}

\begin{proof}
If $x$ is a maximal element, then the total weight of the nodes in $G$ is clearly $|T|$, as every maximal element is added. If $x$ is covered by an element $p$, then the total weight of the nodes in $G$ will be $wt(p)-1$. But $wt(p)>\lceil |T|/2 \rceil$ (as otherwise $p$ would have been returned by \texttt{centralElement} rather than $x$), so we are done.
\end{proof}

\begin{lem}
\label{lem:smallSet}
In each iteration of the loop, $|G|\le 1/\gamma^S$.
\end{lem}

\begin{proof}
We will show that $\gamma^{G}=1/|G|$, implying $\gamma^S \le 1/|G|$. Restrict the marked element to being an element of $G$. Then an algorithm can only remove elements of $G$ from consideration by querying within $G$.
This is because, if $x$ is not a maximal element of $T$, all the members of $G$ are covered by a single element $p$, so the only queries that can allow us to reject members of $G$ are queries to members of $G$.
Alternatively, if $x$ is a maximal element of $T$, then it is easy to see that $x$ is actually also a maximal element of $S$. So $G$ will contain all the maximal elements of $S$, and again the only queries that can allow us to reject members of $G$ are queries to members of $G$.
\end{proof}

\begin{thm}
\label{thm:forestUpperBound}
Algorithm \ref{alg:forest} finds the marked element in a forest-like $n$-element poset $S$, or outputs that no such element exists, with certainty using at most $O\left(\log n/\sqrt{\gamma^S}\right)$ queries to $S$.
\end{thm}

\begin{proof}
It is immediate that the algorithm is correct, as each iteration of the loop is guaranteed to remove at least one element from $T$. It remains to prove an upper bound on its query complexity. If the marked element $a$ is in the set $T$ at all, we are guaranteed that either $a\le x$ for either exactly one element $x \in G$, or for no elements in $G$. The Grover search step will thus either reduce the search space to the elements $\{z\}$ of $T$ for which $z \le x$, or will remove all the elements $z \in T$ that are less than any element in $G$ from consideration. Each element of $G$ has weight at most $\lceil |T|/2 \rceil$, and by Lemma \ref{lem:reduceByHalf}, their total weight is at least $|T|/2$. So each iteration of the loop will reduce the size of $T$ by at least about half. By Lemma \ref{lem:smallSet}, each Grover search uses at most $O(1/\sqrt{\gamma^S})$ queries, so the theorem is proven.
\end{proof}

In some cases, Algorithm \ref{alg:forest} may do better than this upper bound suggests. One such example is searching a completely unstructured set (in which case the algorithm reduces to standard unstructured search, and thus achieves an $O\left(\sqrt{n}\right)=O\left(1/\sqrt{\gamma^S}\right)$ query complexity). As another example, it is easy to convince oneself that Algorithm \ref{alg:forest} finds the marked element in a poset whose Hasse diagram is a complete $k$-ary tree with $l$ levels using $O(\sqrt{k}l)$ queries, rather than the $O(\sqrt{k}l \log k)$ queries guaranteed by Theorem \ref{thm:forestUpperBound}.

Finally, note that the extension to searching for an unknown number of marked elements is straightforward: in this case, the exact Grover search step is replaced by picking an element $y$ from $G$ uniformly at random. If there exists a marked element $a$ such that $a\le y'$ for some element $y'\in G$, then the probability that $y=y'$ is at least $1/\sqrt{\gamma^S}$. We need to boost this success probability to $\Omega(1-1/\log n)$ in order for the success probability after $O(\log n)$ recursions to be $\Omega(1)$. By a result of Buhrman et al \cite{buhrman2} on amplification of classical probabilistic algorithms with one-sided error, this can be achieved using $O(\sqrt{\log \log n}/\sqrt{\gamma^S})$ iterations of picking $y\in G$ uniformly at random, giving an overall complexity of $O\left(\log n \sqrt{\log \log n}/\sqrt{\gamma^S}\right)$.


\section{The concrete model}

In this section, we consider the problem of poset search in the model studied by Linial and Saks \cite{linial}, where the poset is thought of as storing partially sorted integers (or elements from any other totally ordered set), and querying an element of the poset returns the integer stored at that element. Note that we redefine $D(S)$, $Q_E(S)$ and $Q_2(S)$ appropriately.


\subsection{Overall relationships}

This model appears more complex to analyse, as the complexity of the search problem now depends not only on the structure of the poset being searched, but also on the integers that are stored in that poset. Also, the classical analysis of Linial and Saks \cite{linial} relies on certain properties of classical algorithms for poset search that quantum algorithms seem not to share. For example, at the end of a correct classical algorithm which searched unsuccessfully for the element $a$ in $S$, every element $x\in S$ must have been classified according to whether $x<a$, $x=a$ or $x>a$. Quantum algorithms appear not to have this property.

However, we can develop a quantum lower bound that is similar to a known classical lower bound based on the size of the largest ``unsorted'' subset of $S$, namely the size of the largest antichain, $w(S)$. It turns out that finding an element in such a subset reduces to an unstructured search problem. We begin with a lemma whose classical part was shown by Linial and Saks \cite{linial} with a different proof.

\begin{lem}
\label{con:section}
Let $S$ be a poset and let $T$ be a section of $S$. Then $D(S) \ge D(T)$, $Q_E(S) \ge Q_E(T)$ and $Q_2(S) \ge Q_2(T)$.
\end{lem}

\begin{proof}
First, note that $S$ can be partitioned into three disjoint subsets (or {\em layers}): the set $T$; an ``upper'' set $U$ where for all $u \in U$, there is no $t \in T$ such that $u \le t$; and a ``lower'' set $V$ where for all $v \in V$, there is no $t \in T\cup U$ such that $t \le v$. Assume $S$ has $n$ elements, identified with the integers. Let $V$ store the integers $\{1,\dots,|V|\}$ in some manner consistent with its partial order, and similarly let $U$ store the integers $\{|V|+|T|+1,\dots,n\}$. By the definition of the partitioning of $S$, $T$ can store every permutation of the integers $\{|V|+1,\dots,|V|+|T|\}$ that is consistent with its own partial order, independently of the integers stored in the remainder of $S$.

Now consider an adversarial strategy where the marked element is guaranteed to be in the set $\{|V|+1,\dots,|V|+|T|\}$, and thus is stored in $T$. Any query to elements in $U$ or $V$ will then give no information about the position of the marked element within $T$, so any classical or quantum algorithm can restrict itself to making queries to elements in $T$. But any classical [exact quantum, bounded-error quantum] algorithm to find a marked element in $T$ that only makes queries to elements in $T$ must use $D(T)$ [$Q_E(T)$, $Q_2(T)$] queries.
\end{proof}
Note that this property does not hold for arbitrary subsets of posets \cite{linial}: for example, the following posets $S,\,T\subset S$ have $D(S)=3$ but $D(T)=4$. The theorem does not hold at all in the abstract model of poset search discussed in the previous section.
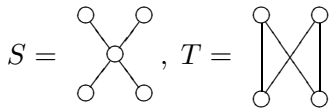
\[
\xymatrix@=5pt{
& \node\edge{rd} & & \node\edge{ld} \\
S = & & \node\edge{ld}\edge{rd} & \\
& \node & & \node}
\xymatrix@=5pt{
& \node\edge{rrdd}\edge{dd} & & \node\edge{lldd}\edge{dd} \\
,\;T = & & & \\
& \node & & \node}
\]
\begin{lem}
\label{con:unordered}
Let $S$ be an $n$-element unstructured poset. Then $D(S) = n$ and $Q_2(S) = \Omega(\sqrt{n})$.
\end{lem}

\begin{proof}
Let $S$ store an arbitrary permutation $\pi$ of the integers $\{1,\dots,n\}$, and let the marked element be $a=\pi(1)$. The classical lower bound is obvious \cite{linial} (as the only information obtained from a query to an element $x \in S$ is whether $a=x$ or $a \neq x$, every element in $S$ may need to be queried in the worst case). In the quantum case, the lower bound of Ambainis on inverting a permutation \cite{ambainis2} may be used to show that any quantum algorithm to find $a$ requires $\Omega(\sqrt{n})$ queries.
\end{proof}

\begin{thm}
\label{con:lower}
Let $S$ be an $n$-element poset. Then $D(S) = \Omega(w(S))$ and $Q_2(S) = \Omega(\sqrt{w(S)})$. Also, $Q_2(S) = \Omega(\log n)$.
\end{thm}

\begin{proof}
Let $T$ be the largest antichain in $S$. $T$ is unstructured, $T$ is a section of $S$ and $|T|=w(S)$. The first part of the theorem follows immediately from Lemma \ref{con:section} and Lemma \ref{con:unordered}. For the second part, note that any quantum algorithm to find a marked element in $S$ could also be used to find a marked element in a totally ordered set of $n$ elements. The lower bound then follows from the lower bound of Ambainis \cite{ambainis3} (improved by H\o yer, Neerbek, and Shi \cite{hoyer}) on quantum search of an ordered list.
\end{proof}
We now consider the question of upper bounds. It turns out that, up to a logarithmic factor, the width $w(S)$ {\em completely} characterises the classical and quantum query complexities of searching in this model. To show this, we will need the following powerful combinatorial result, which says something about the decomposition of a poset into chains.
\begin{thm} {\bf (Dilworth's Theorem \cite{dilworth})}\\
\label{con:dilworth}
Let $S$ be an $n$-element poset with $w(S)=k$. Then $S$ is the union of $k$ disjoint chains.
\end{thm}
In fact, such a decomposition can be found in time $O(n^3)$ \cite{bogart}.
\begin{lem}
\label{con:upper}
Let $S$ be a poset. Then we have $D(S) = O(w(S) \log h(S))$ and $Q_E(S) = O(\sqrt{w(S)} \log h(S))$.
\end{lem}

\begin{proof}
Decompose $S$ into a set $C$ containing $w(S)$ disjoint chains, each of which contains at most $h(S)$ elements. The classical algorithm proceeds by searching each chain in $C$ in turn, using binary search. The total number of queries required is therefore $O(w(S) \log h(S))$.

In the quantum case, our algorithm will nest an exact binary search algorithm within the exact variant of Grover's search algorithm. We produce an oracle $P_a$ which, when given the identifier of a chain in $C$ as input, returns whether the desired element $a$ is contained within that chain; each call to $P_a$ clearly requires at most $O(\log h(S))$ queries to the set. As the chains are disjoint, we are guaranteed that $P_a$ will return 1 on only one input. The exact variant of Grover's algorithm therefore requires (see Theorem \ref{thm:exactGrover}) $O(\sqrt{w(S)})$ queries to $P_a$ to determine which chain (if any) contains $a$. A final $O(\log h(S))$ queries are used to find $a$ within that chain, for an overall query complexity of $O(\sqrt{w(S)} \log h(S))$.
\end{proof}
If the binary search parts of this algorithm are replaced by the use of a quantum ordered search algorithm (e.g.\ \cite{childs}), the query complexity can be improved by a constant factor. Note that this algorithm actually also works in the abstract model of poset search, thus showing that, as one might expect, search in the abstract model is always at least as easy as in the concrete model (up to the $\log h(S)$ factor). Furthermore, note that an extension to search where a given integer may be stored at multiple positions in the poset is immediate: the Grover search steps are replaced by search for an unknown number of marked items \cite{brassard} to give an $O(\sqrt{w(S)} \log h(S))$-query bounded-error quantum algorithm.

We can now show that the classical and quantum query complexities of poset search in the concrete model are polynomially related.
\begin{thm}
Let $S$ be an $n$-element poset with $Q_2(S)=k$. Then $D(S) = O(k^2 \log n) = O(k^3)$.
\end{thm}

\begin{proof}
Follows immediately from the quantum lower bounds of Lemma \ref{con:lower} and the classical upper bound of Lemma \ref{con:upper}.
\end{proof}


\subsection{Searching a partially sorted array}

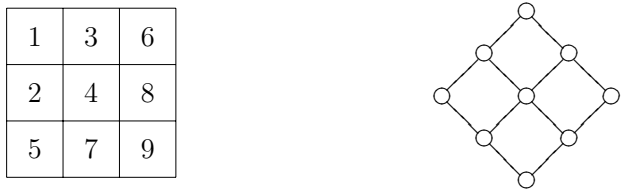
\begin{figure}[htp]
\centering
\subfigure{
\begin{pgfpicture}{4.5cm}{0cm}{2.25cm}{2.25cm}
\begin{pgfmagnify}{0.75}{0.75}
\pgfsetlinewidth{0.4pt}
\pgfgrid[stepx=1cm,stepy=1cm]{\pgfpoint{0cm}{0cm}}{\pgfpoint{3cm}{3cm}}
\pgfputat{\pgfpoint{0.5cm}{0.5cm}}{\pgfbox[center,center]{\Large 5}}
\pgfputat{\pgfpoint{1.5cm}{0.5cm}}{\pgfbox[center,center]{\Large 7}}
\pgfputat{\pgfpoint{2.5cm}{0.5cm}}{\pgfbox[center,center]{\Large 9}}
\pgfputat{\pgfpoint{0.5cm}{1.5cm}}{\pgfbox[center,center]{\Large 2}}
\pgfputat{\pgfpoint{1.5cm}{1.5cm}}{\pgfbox[center,center]{\Large 4}}
\pgfputat{\pgfpoint{2.5cm}{1.5cm}}{\pgfbox[center,center]{\Large 8}}
\pgfputat{\pgfpoint{0.5cm}{2.5cm}}{\pgfbox[center,center]{\Large 1}}
\pgfputat{\pgfpoint{1.5cm}{2.5cm}}{\pgfbox[center,center]{\Large 3}}
\pgfputat{\pgfpoint{2.5cm}{2.5cm}}{\pgfbox[center,center]{\Large 6}}
\end{pgfmagnify}
\end{pgfpicture}
}
\subfigure{
\put(30,60){
$\xymatrix@=10pt{
& & \node\edge{ld}\edge{rd} & & \\
& \node\edge{ld}\edge{rd} & & \node\edge{ld}\edge{rd} & \\
\node\edge{rd} & & \node\edge{ld}\edge{rd} & & \node\edge{ld} \\
& \node\edge{rd} & & \node\edge{ld} & \\
& & \node & &
}$
}
}
\caption{A $3\times3$ 2-dimensional array sorted by rows and columns, and its corresponding Hasse diagram.}
\label{fig:sortedMatrix}
\end{figure}

Consider the following problem. We are given a $d$-dimensional $m_1 \times m_2 \times \cdots \times m_d$ array of integers $T$ that has been sorted in ascending order in each dimension (i.e.\ $(i_1 \le j_1)\,\wedge\,(i_2 \le j_2)\,\wedge \cdots \wedge (i_d \le j_d) \Rightarrow T(i_1, \dots, i_d) \le T(j_1,\dots,j_d)$), and must find a given integer in this array, or output ``not found'', using the minimum number of queries to the array. It is easy to see that this structure gives rise to a partially ordered set; see Figure \ref{fig:sortedMatrix} for the Hasse diagram of such a poset.

We are particularly interested in the special case where $m_i=m$ for all $i$. Call the poset corresponding to such a $d$-dimensional array $S_{d,m}$. Linial and Saks give \cite{linial} an $O(m^{d-1})$ classical algorithm for the problem of searching $S_{d,m}$, which is asymptotically optimal. When $d=2$, it is easy to see that we have $w(S_{2,m})=m$. For higher $d$, Linial and Saks show that $w(S_{d,m})=\Theta(m^{d-1})$. This follows from consideration of the set of elements that are indexed by a position $(i_1,\dots,i_d)$ such that $\sum_k i_k = m+1$; this is clearly an antichain and can be shown to have size $\Theta(m^{d-1})$. It is thus immediate from Lemma \ref{con:upper} and Lemma \ref{con:lower} that there exists a quantum algorithm that searches this poset using $O(m^{(d-1)/2}\,d\log m)$ queries, which is optimal up to the $d\log m$ factor.

We can write down such an algorithm explicitly as follows. The algorithm for $d=1$ is just binary search. For $d=2$, we nest a binary search algorithm on the rows within Grover search on the columns for an overall query complexity of $O(\sqrt{m} \log m)$. For $d=3$, the algorithm simply performs Grover search on $m$ copies of the $d=2$ search algorithm, giving $O(m \log m)$ queries, and so on for $d>3$.

It is worth noting that this poset structure is an example where searching in the abstract model is significantly easier than in the concrete model. Indeed, there exists a simple $O(d \log m)$ classical algorithm for search in the abstract model: simply perform binary search on each dimension of $T$.

In the following section, we will give an asymptotically optimal bounded-error quantum algorithm that searches a 2-dimensional $m \times m$ array of {\em distinct} integers in $O(\sqrt{m})$ queries. This then implies an asymptotically optimal $O(m^{(d-1)/2})$-query algorithm for searching a $d$-dimensional $m \times m \times \cdots \times m$ array of distinct integers. The optimal $d$-dimensional algorithm follows from treating the array as the union of $m^{d-2}$ disjoint 2-dimensional $m \times m$ arrays. Each 2-dimensional array is searched by the optimal algorithm, which is treated as an oracle within an overall application of quantum search. Although the 2-dimensional search algorithm is bounded-error, a version of quantum search which can cope with bounded-error inputs (due to H\o yer, Mosca and de Wolf \cite{hoyer2}) can be used to achieve a constant probability of success in $O(m^{(d-1)/2})$ queries.


\subsubsection{Optimal search of a 2-dimensional array sorted by rows and columns}
\label{sec:2dsearch}

In this section, we give an asymptotically optimal algorithm to search for a known integer $a$ within an $r \times c$ 2-dimensional array of distinct integers sorted by rows and columns. We will start by describing a classical algorithm for the same problem, which is asymptotically (but not exactly \cite{linial}) optimal. The algorithm's operation will be described in terms of the original array, rather than the more abstract poset structure. Call the $\lceil \frac{r}{2} \rceil$'th row of the array the {\em central} row $R$, and similarly let the $\lceil \frac{c}{2} \rceil$'th column be the central column $C$.

If $r\le c$, begin by performing binary search for $a$ on the central column, using $O(\log r)$ queries. If $r > c$, do the same, but on the central row, using $O(\log c)$ queries. Assume $r \le c$ and that $a$ is not in the central column (otherwise, $a$ will be found by the binary search, and can be returned immediately). Then by the end of the binary search we will have found an element $x$ such that $x=\max_{x'\in C}(x' < a)$, and an element $y$ such that $y=\min_{y'\in C}(y' > a)$ (so $y$ is positioned directly below $x$ in the array). This then implies that all elements in the array above and to the left of $x$ are also less than $a$, and similarly all elements below and to the right of $y$ are greater than $a$, so these elements can be discarded. As $x$ and $y$ are in the central column, we must have excluded at least half of the elements in the array from consideration.

We are then left with two smaller instances of the same problem to solve: the subarray below and to the left of $y$, and the subarray above and to the right of $x$. The algorithm proceeds to search these subarrays recursively until $a$ is found, performing binary search on central rows or central columns as appropriate.

\begin{figure}[htp]
\centering
\subfigure{
\begin{pgfpicture}{0cm}{0cm}{2.5cm}{2.5cm}
\begin{pgfmagnify}{0.5}{0.5}
\color{lightgray!70}
\color{darkgray!70}
\pgfrect[fill]{\pgfpoint{2cm}{0cm}}{\pgfpoint{1cm}{5cm}}
\color{black}
\pgfsetlinewidth{0.4pt}
\pgfgrid[stepx=1cm,stepy=1cm]{\pgfpoint{0cm}{0cm}}{\pgfpoint{5cm}{5cm}}
\pgfputat{\pgfpoint{0.5cm}{4.5cm}}{\pgfbox[center,center]{\LARGE 1}}
\pgfputat{\pgfpoint{1.5cm}{4.5cm}}{\pgfbox[center,center]{\LARGE 3}}
\pgfputat{\pgfpoint{2.5cm}{4.5cm}}{\pgfbox[center,center]{\LARGE 5}}
\pgfputat{\pgfpoint{3.5cm}{4.5cm}}{\pgfbox[center,center]{\LARGE 10}}
\pgfputat{\pgfpoint{4.5cm}{4.5cm}}{\pgfbox[center,center]{\LARGE 13}}

\pgfputat{\pgfpoint{0.5cm}{3.5cm}}{\pgfbox[center,center]{\LARGE 2}}
\pgfputat{\pgfpoint{1.5cm}{3.5cm}}{\pgfbox[center,center]{\LARGE 4}}
\pgfputat{\pgfpoint{2.5cm}{3.5cm}}{\pgfbox[center,center]{\LARGE 7}}
\pgfputat{\pgfpoint{3.5cm}{3.5cm}}{\pgfbox[center,center]{\LARGE \bf 11}}
\pgfputat{\pgfpoint{4.5cm}{3.5cm}}{\pgfbox[center,center]{\LARGE 14}}

\pgfputat{\pgfpoint{0.5cm}{2.5cm}}{\pgfbox[center,center]{\LARGE 6}}
\pgfputat{\pgfpoint{1.5cm}{2.5cm}}{\pgfbox[center,center]{\LARGE 8}}
\pgfputat{\pgfpoint{2.5cm}{2.5cm}}{\pgfbox[center,center]{\LARGE 9}}
\pgfputat{\pgfpoint{3.5cm}{2.5cm}}{\pgfbox[center,center]{\LARGE 15}}
\pgfputat{\pgfpoint{4.5cm}{2.5cm}}{\pgfbox[center,center]{\LARGE 21}}

\pgfputat{\pgfpoint{0.5cm}{1.5cm}}{\pgfbox[center,center]{\LARGE 12}}
\pgfputat{\pgfpoint{1.5cm}{1.5cm}}{\pgfbox[center,center]{\LARGE 16}}
\pgfputat{\pgfpoint{2.5cm}{1.5cm}}{\pgfbox[center,center]{\LARGE 17}}
\pgfputat{\pgfpoint{3.5cm}{1.5cm}}{\pgfbox[center,center]{\LARGE 20}}
\pgfputat{\pgfpoint{4.5cm}{1.5cm}}{\pgfbox[center,center]{\LARGE 24}}

\pgfputat{\pgfpoint{0.5cm}{0.5cm}}{\pgfbox[center,center]{\LARGE 18}}
\pgfputat{\pgfpoint{1.5cm}{0.5cm}}{\pgfbox[center,center]{\LARGE 19}}
\pgfputat{\pgfpoint{2.5cm}{0.5cm}}{\pgfbox[center,center]{\LARGE 22}}
\pgfputat{\pgfpoint{3.5cm}{0.5cm}}{\pgfbox[center,center]{\LARGE 23}}
\pgfputat{\pgfpoint{4.5cm}{0.5cm}}{\pgfbox[center,center]{\LARGE 25}}
\end{pgfmagnify}
\end{pgfpicture}
}
\subfigure{
\begin{pgfpicture}{0cm}{0cm}{2.5cm}{2.5cm}
\begin{pgfmagnify}{0.5}{0.5}
\color{lightgray!70}
\pgfrect[fill]{\pgfpoint{0cm}{2cm}}{\pgfpoint{3cm}{3cm}}
\pgfrect[fill]{\pgfpoint{2cm}{0cm}}{\pgfpoint{3cm}{2cm}}
\color{darkgray!70}
\pgfrect[fill]{\pgfpoint{1cm}{0cm}}{\pgfpoint{1cm}{2cm}}
\pgfrect[fill]{\pgfpoint{3cm}{3cm}}{\pgfpoint{2cm}{1cm}}
\color{black}
\pgfsetlinewidth{1.2pt}
\pgfrect[stroke]{\pgfpoint{3cm}{2cm}}{\pgfpoint{2cm}{3cm}}
\pgfrect[stroke]{\pgfpoint{0cm}{0cm}}{\pgfpoint{2cm}{2cm}}
\pgfsetlinewidth{0.4pt}
\pgfgrid[stepx=1cm,stepy=1cm]{\pgfpoint{0cm}{0cm}}{\pgfpoint{5cm}{5cm}}
\pgfputat{\pgfpoint{0.5cm}{4.5cm}}{\pgfbox[center,center]{\LARGE 1}}
\pgfputat{\pgfpoint{1.5cm}{4.5cm}}{\pgfbox[center,center]{\LARGE 3}}
\pgfputat{\pgfpoint{2.5cm}{4.5cm}}{\pgfbox[center,center]{\LARGE 5}}
\pgfputat{\pgfpoint{3.5cm}{4.5cm}}{\pgfbox[center,center]{\LARGE 10}}
\pgfputat{\pgfpoint{4.5cm}{4.5cm}}{\pgfbox[center,center]{\LARGE 13}}

\pgfputat{\pgfpoint{0.5cm}{3.5cm}}{\pgfbox[center,center]{\LARGE 2}}
\pgfputat{\pgfpoint{1.5cm}{3.5cm}}{\pgfbox[center,center]{\LARGE 4}}
\pgfputat{\pgfpoint{2.5cm}{3.5cm}}{\pgfbox[center,center]{\LARGE 7}}
\pgfputat{\pgfpoint{3.5cm}{3.5cm}}{\pgfbox[center,center]{\LARGE \bf 11}}
\pgfputat{\pgfpoint{4.5cm}{3.5cm}}{\pgfbox[center,center]{\LARGE 14}}

\pgfputat{\pgfpoint{0.5cm}{2.5cm}}{\pgfbox[center,center]{\LARGE 6}}
\pgfputat{\pgfpoint{1.5cm}{2.5cm}}{\pgfbox[center,center]{\LARGE 8}}
\pgfputat{\pgfpoint{2.5cm}{2.5cm}}{\pgfbox[center,center]{\LARGE 9}}
\pgfputat{\pgfpoint{3.5cm}{2.5cm}}{\pgfbox[center,center]{\LARGE 15}}
\pgfputat{\pgfpoint{4.5cm}{2.5cm}}{\pgfbox[center,center]{\LARGE 21}}

\pgfputat{\pgfpoint{0.5cm}{1.5cm}}{\pgfbox[center,center]{\LARGE 12}}
\pgfputat{\pgfpoint{1.5cm}{1.5cm}}{\pgfbox[center,center]{\LARGE 16}}
\pgfputat{\pgfpoint{2.5cm}{1.5cm}}{\pgfbox[center,center]{\LARGE 17}}
\pgfputat{\pgfpoint{3.5cm}{1.5cm}}{\pgfbox[center,center]{\LARGE 20}}
\pgfputat{\pgfpoint{4.5cm}{1.5cm}}{\pgfbox[center,center]{\LARGE 24}}

\pgfputat{\pgfpoint{0.5cm}{0.5cm}}{\pgfbox[center,center]{\LARGE 18}}
\pgfputat{\pgfpoint{1.5cm}{0.5cm}}{\pgfbox[center,center]{\LARGE 19}}
\pgfputat{\pgfpoint{2.5cm}{0.5cm}}{\pgfbox[center,center]{\LARGE 22}}
\pgfputat{\pgfpoint{3.5cm}{0.5cm}}{\pgfbox[center,center]{\LARGE 23}}
\pgfputat{\pgfpoint{4.5cm}{0.5cm}}{\pgfbox[center,center]{\LARGE 25}}
\end{pgfmagnify}
\end{pgfpicture}
}
\subfigure{
\begin{pgfpicture}{0cm}{0cm}{2.5cm}{2.5cm}
\begin{pgfmagnify}{0.5}{0.5}
\color{lightgray!70}
\pgfrect[fill]{\pgfpoint{0cm}{0cm}}{\pgfpoint{5cm}{5cm}}
\color{white}
\pgfrect[fill]{\pgfpoint{3cm}{3cm}}{\pgfpoint{1cm}{1cm}}
\color{black}
\pgfsetlinewidth{1.2pt}
\pgfrect[stroke]{\pgfpoint{3cm}{3cm}}{\pgfpoint{1cm}{1cm}}
\pgfsetlinewidth{0.4pt}
\pgfgrid[stepx=1cm,stepy=1cm]{\pgfpoint{0cm}{0cm}}{\pgfpoint{5cm}{5cm}}
\pgfputat{\pgfpoint{0.5cm}{4.5cm}}{\pgfbox[center,center]{\LARGE 1}}
\pgfputat{\pgfpoint{1.5cm}{4.5cm}}{\pgfbox[center,center]{\LARGE 3}}
\pgfputat{\pgfpoint{2.5cm}{4.5cm}}{\pgfbox[center,center]{\LARGE 5}}
\pgfputat{\pgfpoint{3.5cm}{4.5cm}}{\pgfbox[center,center]{\LARGE 10}}
\pgfputat{\pgfpoint{4.5cm}{4.5cm}}{\pgfbox[center,center]{\LARGE 13}}

\pgfputat{\pgfpoint{0.5cm}{3.5cm}}{\pgfbox[center,center]{\LARGE 2}}
\pgfputat{\pgfpoint{1.5cm}{3.5cm}}{\pgfbox[center,center]{\LARGE 4}}
\pgfputat{\pgfpoint{2.5cm}{3.5cm}}{\pgfbox[center,center]{\LARGE 7}}
\pgfputat{\pgfpoint{3.5cm}{3.5cm}}{\pgfbox[center,center]{\LARGE \bf 11}}
\pgfputat{\pgfpoint{4.5cm}{3.5cm}}{\pgfbox[center,center]{\LARGE 14}}

\pgfputat{\pgfpoint{0.5cm}{2.5cm}}{\pgfbox[center,center]{\LARGE 6}}
\pgfputat{\pgfpoint{1.5cm}{2.5cm}}{\pgfbox[center,center]{\LARGE 8}}
\pgfputat{\pgfpoint{2.5cm}{2.5cm}}{\pgfbox[center,center]{\LARGE 9}}
\pgfputat{\pgfpoint{3.5cm}{2.5cm}}{\pgfbox[center,center]{\LARGE 15}}
\pgfputat{\pgfpoint{4.5cm}{2.5cm}}{\pgfbox[center,center]{\LARGE 21}}

\pgfputat{\pgfpoint{0.5cm}{1.5cm}}{\pgfbox[center,center]{\LARGE 12}}
\pgfputat{\pgfpoint{1.5cm}{1.5cm}}{\pgfbox[center,center]{\LARGE 16}}
\pgfputat{\pgfpoint{2.5cm}{1.5cm}}{\pgfbox[center,center]{\LARGE 17}}
\pgfputat{\pgfpoint{3.5cm}{1.5cm}}{\pgfbox[center,center]{\LARGE 20}}
\pgfputat{\pgfpoint{4.5cm}{1.5cm}}{\pgfbox[center,center]{\LARGE 24}}

\pgfputat{\pgfpoint{0.5cm}{0.5cm}}{\pgfbox[center,center]{\LARGE 18}}
\pgfputat{\pgfpoint{1.5cm}{0.5cm}}{\pgfbox[center,center]{\LARGE 19}}
\pgfputat{\pgfpoint{2.5cm}{0.5cm}}{\pgfbox[center,center]{\LARGE 22}}
\pgfputat{\pgfpoint{3.5cm}{0.5cm}}{\pgfbox[center,center]{\LARGE 23}}
\pgfputat{\pgfpoint{4.5cm}{0.5cm}}{\pgfbox[center,center]{\LARGE 25}}
\end{pgfmagnify}
\end{pgfpicture}
}
\caption{Example of the classical algorithm's operation when searching for the element 11: dark grey squares are those that are searched in each round, light grey squares have been excluded from consideration, white squares are still to be searched. Here, 11 is found with only 2 levels of recursion.}
\end{figure}

How many queries to the array does this algorithm require? Let $T(m)$ denote the number of queries used to search an $r \times c$ array, with $m=\max(r,c)$. Then it is easy to see that $T(m)$ will be maximised if each level of binary search always terminates as close to the centre of the central column/row as possible (thus maximising the number of queries required for binary search in the next level of recursion). We therefore have
\be T(m) \le \lceil \log_2 m+1 \rceil + 2 T(m/2) \ee
and unwinding the recursion gives $T(m) = O(m)$.

We would like to find an analogous quantum algorithm that achieves some reduction in queries by searching the subarrays in superposition, rather than sequentially. In fact, it turns out that we can make a general statement about when recursive classical search algorithms can be turned into improved quantum search algorithms, which is given as the following lemma. The proof is a fairly straightforward generalisation of a powerful result of Aaronson and Ambainis \cite{aaronson2}, so is deferred to Appendix \ref{appendixA}.

\begin{lem}
\label{lem:stateRecurse}
Let $P_n$ be the problem of searching an abstract database, parametrised by an abstract size $n$, for a known element which may or may not be in the database. Let $T(n)$ be the time required for a bounded-error quantum algorithm to solve $P_n$, i.e.\ to find the element, or output ``not found''. Let $P_n$ satisfy the following conditions:
\begin{itemize}
\item If $n\le n_0$ for some constant $n_0$, then there exists an algorithm to find the element, if it is contained in the database, in time $T(n) \le t_0$, for some constant $t_0$.
\item If $n>n_0$, then the database can be divided into $k$ sub-databases of size at most $\lceil n/k \rceil$, for some constant $k>1$.
\item If the element is contained in the original database, then it is contained in exactly one of these sub-databases.
\item Each division into sub-databases uses time $f(n)$, where $f(n)=O(n^{1/2-\epsilon})$ for some $\epsilon>0$.
\end{itemize}
Then $T(n)=O(\sqrt{n})$.
\end{lem}

We show that the search problem in question fits the conditions of the lemma. We consider the problem to be parametrised by a ``size'' $m=\max(r,c)$. Assuming that $a$ is stored in the set and is not stored in the central row/column, one step of the classical procedure given above will divide any array of size $m$ into two arrays of size at most $\lceil m/2 \rceil$, exactly one of which contains $a$, in time $O(\log m)$. This division can be performed recursively until the arrays are reduced to a constant size. In the case where the binary search of the central row/column actually finds $a$, the algorithm can easily be modified to not return $a$ immediately, but to restrict the search area in the next recursion to two subarrays, exactly one of which includes $a$, and both of which are of size at most $\lceil m/2 \rceil$.

There thus exists a quantum algorithm, given explicitly in Appendix \ref{appendixA}, that can find an arbitrary element $a$ in the array in $O(\sqrt{m})$ time, and hence $O(\sqrt{m})$ queries.


\subsubsection{Finding the intersection of two increasing lists}
\label{sec:intersect}

Classically, there is a correspondence between the problem of searching in an $r \times c$ array sorted by rows and columns and merging two sorted lists of length $r$ and $c$: any decision tree for the one problem gives a decision tree for the other \cite{linial}. However, this does not appear to hold for quantum algorithms; indeed, it is straightforward to show, using Holevo's Theorem \cite{holevo}, an $\Omega(r+c)$ quantum query lower bound for the merge problem. Nevertheless, we can define a natural search problem that turns out to arise from the poset search problem.

{\bf Problem:} Given two lists of integers in increasing order, output an integer that occurs in both lists, or report that no such integer exists.

This can be thought of as a special case of the element distinctness problem \cite{aaronson}. It was studied by Buhrman et al \cite{buhrman}, who also refer to it as the {\em monotone claw} problem (a claw is an input on which two functions take the same value). Let the lists be denoted $L$ and $M$ and be of length $l$ and $m$ respectively, with $l\ge m$. Then the ingenious algorithm of \cite{buhrman} finds an integer occuring in both lists using $O(\sqrt{l} c^{\log^* l})$ queries, where $\log^*$ is the iterated logarithm function and $c$ is a constant. This algorithm can easily be translated into the setting of poset search, and allows an $m \times m$ array that is sorted by rows and columns, and may contain duplicates, to be searched using $O(\sqrt{m} c^{\log^* m})$ time for some constant $c$.

Here, we will go in the other direction, and show that the algorithm of Section \ref{sec:2dsearch} can be used to find the integer occurring in both sorted lists using $O(\sqrt{l})$ time. As noted in \cite{buhrman}, there is an $\Omega(\sqrt{l})$ lower bound for this problem, so the algorithm given here is asymptotically optimal. However, as $c^{\log^* l}$ is already a near-constant function, the new algorithm may be only of theoretical interest, and we describe it briefly.

Consider a notional $l \times m$ array $T$ where entry $T(x,y)$ contains the value $L_x - M_{m+1-y}$. Querying one entry of $T$ uses one query to each list. As the entries in $L$ and $M$ are in increasing order, it is easy to see that $T$ is increasing along rows and columns, and that finding a 0 entry in $T$ corresponds to finding an element of $L$ that also occurs in $M$. Call such an element a {\em match}. If there is only one match, it is immediate that the algorithm of the previous section can be used to find the single 0 entry in $T$, or output that no such entry exists, in time $O(\sqrt{l})$.

There are two possible reasons for there being more than one match. The first is that $L$ and $M$ may contain duplicate elements (i.e.\ may be increasing but not strictly increasing). If this is the case, and if one of the duplicate elements in $L$ (say) is also in $M$, there will be a contiguous rectangle of 0 entries in the array $T$ (call this a {\em zero block}), rather than a single 0. Assume that there is only one zero block. Then the algorithm of Section \ref{sec:2dsearch} must be modified to ensure that, after any splitting of the array into two subarrays, at most one of these arrays contains a 0 entry; i.e.\ to ensure that the zero block does not get split across subarrays. This is necessary to ensure that the conditions of Lemma \ref{lem:stateRecurse} are satisfied. It is easy to see that, in each round of recursion, the zero block can only be split if it lies across a row or column that is used for binary search in that recursion. In order to ensure that only one of the two subarrays produced contains part of the zero block in this case, the binary search of a row (column) can simply be modified to split on the first or last zero entry in that row (column), with no change to the asymptotic complexity. Call this new algorithm the single-block algorithm.

The second case where there may be more than one match is when there is more than one element in $L$ that also occurs in $M$ (or vice versa). In this case, the idea (inspired by \cite{aaronson2}) is to reduce the problem to searching for a single zero block by probabilistically removing elements from the lists. The extended algorithm first runs the single-block algorithm. Assuming that this algorithm outputs ``not found'', the next step is to produce a new pair of smaller lists $L^{(2)}$ and $M^{(2)}$, which will give rise to a notional array $T^{(2)}$, where $T^{(2)}(x,y)=L^{(2)}_x - M^{(2)}_{m+1-y}$.

The reduction in size is achieved by first splitting each list into chunks of size 2. One element (picked at random) within each chunk of $L$ is included in $L^{(2)}$, and similarly for $M$ and $M^{(2)}$. The single-block algorithm is then run on these smaller lists. Assuming that the result is again ``not found'', the chunk size is doubled to 4, and the process repeats, using a chunk size of $2^k$ in each round $k$. Assuming that the single-block algorithm does not find a match in any of the $O(\log l)$ rounds, the final output is ``not found''. The time required for this overall algorithm is then bounded by $O\left(\sum_k \sqrt{l/2^k}\right)=O(\sqrt{l})$.

We sketch a proof that this algorithm succeeds with constant probability. First, it is easy to see that there can be at most one zero block in each row and column of the array $T^{(k)}$ in any round $k$. Using this, one can show that, if there are $z$ zero blocks in $T$, the probability that exactly one remains in $T^{(k)}$ is at least $z/2^{2k}(1-z/2^{2k})$. If we take $k=\lceil \log z/2 \rceil + 1$, this is lower bounded by a constant, so for any $z$ the single-block algorithm succeeds with constant probability in at least one round.


\section{Conclusions}
\label{sec:conclusions}

We have given general upper and lower bounds on quantum search of partially ordered sets, in two different models. Satisfyingly, in the two cases where results were already known on poset search (i.e.\ totally ordered sets and unstructured sets), our lower bounds reduce to known lower bounds, and our new quantum algorithms are (asymptotically) as efficient as the known most efficient algorithms. The bounds in the concrete model are perhaps particularly interesting, because they follow from decomposing a poset into ``structured'' and ``unstructured'' components, and show that, intuitively, almost all the speed-up that can be obtained from quantum search of a poset $S$ is obtained from searching the unstructured parts of $S$.

Although we concentrated on the model of query complexity, our quantum algorithms in both models are efficient in the sense that, given a poset $S$ to be searched, quantum circuits for the algorithms given here can be produced in time polynomial in the size of $S$. Also, the non-query transformations used by the algorithms given here are efficiently implementable.

However, there are still several open questions. Firstly: in the abstract model, is there a general lower bound of $Q(S) = \Omega(\log n)$? This would be an interesting generalisation of the known logarithmic quantum lower bound on searching an ordered list \cite{ambainis3,hoyer}. Also, can the logarithmic factors in the quantum upper bounds in both models be improved, perhaps by being changed into additive terms?

There are several possible extensions involving search for multiple marked elements. In the abstract model, can a $O\left(\log n/\sqrt{\gamma^S}\right)$-query algorithm be produced for search for multiple marked elements in arbitrary posets? In the concrete model, could the algorithm of section \ref{sec:2dsearch} be extended to arrays that may contain duplicate elements?


\section*{Acknowledgements}

I would like to thank Richard Jozsa, Rapha\"el Clifford, Richard Low, Dan Shepherd and Aram Harrow for helpful discussions.


\appendix

\section{Amplitude amplification of recursive search}
\label{appendixA}

The aim of this appendix is to give a proof of a somewhat generalised version of a powerful result that was shown by Aaronson and Ambainis \cite{aaronson2} in the course of their work on quantum search of spatial regions. Informally, we would like to be able to find ``cookbook'' quantum algorithms for search problems for which there exists a recursive classical algorithm. We imagine that we are searching for a distinguished element in an abstract ``database'' that is parametrised by an abstract ``size'' $n$, which is some function of the number of elements in the database. We also imagine that we have the ability to search the database recursively: that is, in time given by some function $f(n)$, we can reduce the search problem to searching $k$ instances of databases of size $\le \lceil n/k \rceil$, for some constant $k>1$.

It is straightforward to show that, classically, the marked element can be found deterministically in $O(n)$ time, by repeated use of this recursive search. An alternative probabilistic classical algorithm for this problem would be: split the input into a number of parts, pick one part uniformly at random, and call yourself recursively on that part. Our quantum algorithm will apply amplitude amplification to this probabilistic algorithm. It will turn out to be advantageous to only amplify a small number of times within the recursive algorithm, and then to amplify again at the end. Amplifying to high probabilities too soon is less efficient \cite{aaronson2}; conversely, if amplitude amplification were only applied at the end of the algorithm, we would require $\Omega(\sqrt{n})$ iterations to amplify the probability to a constant. If the process of dividing the input required time $f(n)=\omega(1)$, this would hurt the overall complexity.

The fundamental amplitude amplification result of Brassard et al \cite{brassard} states that, given a quantum algorithm $A$ with success probability $\epsilon$, we can achieve a success probability of $\Omega(1)$ with only $O(1/\sqrt{\epsilon})$ uses of $A$. However, here we will need a tighter analysis due to Aaronson and Ambainis \cite{aaronson2}, as constants are important within the recursive algorithm.

\begin{lem}
\label{lem:ampamp}
Given a quantum algorithm with success probability at least $\epsilon$, then by executing it $t=2m+1$ times, where $m \le \pi/(\arcsin \sqrt{\epsilon})-1/2$, we can achieve success probability at least $(1-\frac13 t^2 \epsilon)t^2\epsilon$.
\end{lem}

We are now ready to give a formal definition of a quantum algorithm for recursive search problems, and to upper-bound its time complexity. The algorithm and its analysis closely follow the results on spatial search of a $d$-dimensional cube of \cite{aaronson2}.

\begin{lem}
\label{lem:master}
Let $P_n$ be the problem of searching an abstract database, parametrised by an abstract size $n$, for a known element which may or may not be in the database. Let $T(n)$ be the time required for a bounded-error quantum algorithm to solve $P_n$, i.e.\ to find the element, or output ``not found''. Let $P_n$ satisfy the following conditions:
\begin{itemize}
\item If $n\le n_0$ for some constant $n_0$, then there exists an algorithm to find the element, if it is contained in the database, in time $T(n) \le t_0$, for some constant $t_0$.
\item If $n>n_0$, then the database can be divided into $k$ sub-databases of size at most $\lceil n/k \rceil$, for some constant $k>1$.
\item If the element is contained in the original database, then it is contained in exactly one of these sub-databases.
\item Each division into sub-databases uses time $f(n)$, where $f(n)=O(n^{1/2-\epsilon})$ for some $\epsilon>0$.
\end{itemize}
Then $T(n)=O(\sqrt{n})$.
\end{lem}

\begin{proof}
Our quantum algorithm will be parametrised by two constants $\alpha$ and $\delta$, whose values we will take to be $\delta=\epsilon/2$, $\alpha=\frac{\epsilon(4-3\epsilon)}{8(2-\epsilon)}$, and will be based on the following probabilistic classical algorithm:

If $n\le n_0$, then find the desired element directly or output ``not found'' (using at most $t_0$ steps). Otherwise, assume that there exists an integer $l$ such that $n^\delta = k^l$ \footnote{We assume here that $l$ and $n^\alpha$ are integers. One can show that the need to round these quantities up or down has no effect on the overall asymptotic complexity.}. Recursively divide the problem into subproblems $l$ times, leaving $n^\delta$ subproblems, each of size at most $n^{1-\delta}$. Pick one of the parts at random, and call yourself recursively on that part. Repeat until the desired element has been found.

We will perform a number of iterations of amplitude amplification on this algorithm such that it is executed $n^\alpha$ times. Then we have
\bea
T(n) &\le& n^\alpha \left(\sum_{i=0}^{l-1} k^i f(n/k^i)+T(n^{1-\delta}) \right)\\
&\le& n^\alpha \left(l n^\delta f(n)+T(n^{1-\delta}) \right)\\
&=& n^\alpha f'(n)+n^{\alpha(1+(1-\delta))}f'(n^{1-\delta})+n^{\alpha(1+(1-\delta)+(1-\delta)^2)}f'(n^{(1-\delta)^2})+...+t_0\\
&=& O(n^{\alpha(1+(1-\delta)+(1-\delta)^2+...)})\\
&=& O(n^{\alpha/\delta})
\eea
where we define $f'(n) = l n^\delta f(n) = O(n^{(1-\epsilon)/2} \log n)$. The fourth line follows because $(1-\epsilon)/2 < \alpha(1/\delta-1)$, so for any $m \ge 0$ we have $f'(n^{(1-\delta)^m}) = O(n^{(1-\delta)^m(1-\epsilon)/2} \log n) = o(n^{(\alpha/\delta)(1-\delta)^{m+1}})$, so the $f'(n^{(1-\delta)^m})$ parts of the third line are negligible.

We now calculate a lower bound on the probability of success $P(n)$ of this algorithm. If there were no amplification, we would have $P(n) \ge n^{-\delta} P(n^{1-\delta})$ for $n>n_0$, and $P(n)=1$ for $n\le n_0$. So, by Lemma \ref{lem:ampamp}, we have
\bea
P(n) &\ge& (1-n^{2\alpha-\delta}/3)n^{2\alpha-\delta}P(n^{1-\delta})\\
&=& [(1-n^{2\alpha-\delta}/3)(1-n^{(2\alpha-\delta)(1-\delta)}/3)\cdots]\,n^{(2\alpha-\delta)(1+(1-\delta)+(1-\delta)^2+...)}\\
&=& [(1-n^{2\alpha-\delta}/3)(1-n^{(2\alpha-\delta)(1-\delta)}/3)\cdots]\, \Omega(n^{2\alpha/\delta-1})
\eea
We claim that the remaining product of bracketed terms is lower bounded by a constant that does not depend on $n$. First, note that the algorithm recurses $R$ times, for some $R=O(\log \log n)$. Now
\be
\prod_{k=0}^R (1-\frac13 n^{(2\alpha-\delta)(1-\delta)^k}) \ge 1-\frac13 \sum_{k=0}^{O(\log \log n)} n^{(2\alpha-\delta)(1-\delta)^k} \ge 1-O(n^{2\alpha-\delta}\log \log n) = 1-o(1) \ee
giving the result $P(n)=\Omega(n^{2\alpha/\delta-1})$.

By wrapping this algorithm in another level of amplitude amplification, we can use $O(P(n)^{-1/2})$ iterations of it to achieve a constant probability of success of finding the marked element in time $O(T(n)P(n)^{-1/2})=O(n^{\alpha/\delta}n^{1/2-\alpha/\delta})=O(\sqrt{n})$.
\end{proof}


\end{document}